\documentclass[aps,prd,twocolumn,eqsecnum,showpacs,amsmath]{revtex4}
\usepackage[dvips]{color,graphicx}
\usepackage{amsfonts}
\usepackage{amssymb}
\usepackage{latexsym}

\newcommand{\ma}[1]{\mbox{$\mathcal{#1}$}}

\newcommand{\dalm}{\kern1pt\vbox{\hrule height 0.9pt\hbox{\vrule width
0.9pt\hskip 2.5pt\vbox{\vskip 5.5pt}\hskip 3pt\vrule width
0.3pt}\hrule height 0.3pt}\kern1pt}
\newcommand{\lw}[1]{\smash{\lower2.ex\hbox{#1}}}

\def \N{I\!\!N}
\begin{document}


\title{Lovelock black holes with a nonlinear Maxwell field}
\author{Hideki Maeda$^{1}$}
\email{hideki-at-cecs.cl}
\author{Mokhtar Hassa\"{\i}ne$^{2}$}
\email{hassaine-at-inst-mat.utalca.cl}
\author{Cristi{\'a}n Mart\'{\i}nez$^{1,3}$}
\email{martinez-at-cecs.cl}

\address{
$^{1}$Centro de Estudios Cient\'{\i}ficos (CECS),  Casilla 1469,
Valdivia,
Chile\\
$^{2}$Instituto de Matem\'atica y F\'{\i}sica, Universidad de Talca,
Casilla 747, Talca, Chile\\
$^{3}$Centro de Ingenier\'{\i}a de la Innovaci\'on del CECS (CIN),
Valdivia, Chile}

\date{\today}

\begin{abstract}
We derive electrically charged black hole solutions of the
Einstein-Gauss-Bonnet equations with a nonlinear electrodynamics
source in $n (\ge 5)$ dimensions. The spacetimes are given as a warped product ${\ma M}^{2}
\times {\ma K}^{n-2}$, where ${\ma K}^{n-2}$ is a
$(n-2)$-dimensional constant curvature space. We establish a
generalized Birkhoff's theorem by showing that it is the unique
electrically charged solution with this isometry and for which the
orbit of the warp factor on ${\ma K}^{n-2}$ is non-null. An
extension of the analysis for full Lovelock gravity is also achieved
with a particular attention to the Chern-Simons case.
\end{abstract}

\pacs{
04.20.Jb 
04.40.Nr 
04.50.-h, 
04.50.Gh 
04.50.Kd 
}


\maketitle

\section{Introduction}
Gravitation physics in higher dimensions has been recently
investigated in a focused way mainly motivated by string theory.
Higher-dimensional general relativity is realized in the lowest
order of the Regge slope expansion of strings. Even in general
relativity, black holes in higher dimensions have much richer
structures than those in four dimensions~\cite{hdbh}. The next
stringy correction yields the quadratic Riemann curvature terms in
the heterotic string case~\cite{Gross,gs1987}. In order that the graviton
amplitude is ghost-free, a special combination of the remaining
curvature-squared terms is required yielding to the renormalizable
Gauss-Bonnet term~\cite{Zwiebach:1985uq}.

The origin of considering higher-order curvature invariants lies in
the attempt of generalizing the theory of general relativity in
higher dimensions. Indeed, under the standard assumptions of general
relativity it is natural to describe the spacetime geometry in three
and four dimensions by the Einstein-Hilbert action while for
dimensions greater than four, a more general theory is available.
This fact has been first noticed by Lanczos~\cite {LAN} in five
dimensions and later generalized by Lovelock~\cite{LOV} for
arbitrary dimensions $n$.

The resulting theory is described by the so-called Lovelock
Lagrangian which is a $n$-form constructed with the vielbein $e^a$,
the spin connection $\omega^{ab}$, and their exterior derivatives
without using the Hodge dual. The Lagrangian is a polynomial of
degree $[n/2]$ in the curvature two-form, $R^{ab} = d\,\omega^{ab} +
\omega^{a}_{\;c} \wedge \omega^{cb}$, given by
\begin{equation}
{\mathcal L}^{(n)} = \sum_{p=0}^{[\frac{n-1}{2}]}\alpha_p~
\epsilon_{a_1\cdots a_n} R^{a_1a_2}\cdots
R^{a_{2p-1}a_{2p}}e^{a_{2p+1}}\cdots  e^{a_n} ~, \label{LovLag}
\end{equation}
where [$x$] denotes the integer part of $x$, $\alpha_p$ being arbitrary dimensionful coupling constants and
wedge products between forms are understood. The corresponding
action contains the same degrees of freedom as the Einstein-Hilbert
action~\cite{TEZ}.

The local Lorentz invariance of the Lovelock action (\ref{LovLag})
can be extended into a local (anti-)de~Sitter ((A)dS) symmetry in
odd dimensions by fixing properly the Lovelock coefficients $\alpha_p$. For the AdS
case the coefficients are given by
\begin{equation}
\alpha_p=\frac{1}{n-2p}\left(
\begin{array}{c}
 [\frac{n-1}{2}] \\
 p
 \end{array}
 \right), \label{alphapads}
\end{equation}
where the AdS radius was set equal to 1.
The resulting
Lagrangian belongs to the class of Chern-Simons gauge theories with
Yang-Mills gauge symmetries, and admit supersymmetric extensions.
(See~\cite{Zanelli:2005sa} and references therein.)

It is clear that these higher-curvature terms come into play in
extremely curved regions. Black holes and singularities are one of
the best testbeds for demonstrating the effects of these
higher-curvature terms. There exists an extensive literature about
the exact black-hole solutions, the thermodynamics, the stability,
and other topics concerning the Gauss-Bonnet or more generally the
Lovelock theory. (See~\cite{Charmousis:2008kc,Garraffo:2008hu} for
detailed recent reviews on the subject.)

In the present paper, we shall consider the Gauss-Bonnet and more
generally the Lovelock action in presence of a nonlinear
electrodynamics source given as an arbitrary power $q$ of the
Maxwell invariant,
\begin{eqnarray}
\int d^nx\sqrt{-g}(F_{\mu\nu}F^{\mu\nu})^q. \label{nes}
\end{eqnarray}
Not being exactly the same form as above, the higher $F$-terms also appear in the low-energy limit of heterotic string theory~\cite{gs1987}.
The Gauss-Bonnet black holes with the higher $F$-terms have been investigated in~\cite{GBF}.
Our action (\ref{nes}) may be considered as the simplest model of such higher $F$-terms.

The nonlinear source (\ref{nes}) has been considered in general
relativity~\cite{hm2008} where it has been derived black-hole
solutions with interesting asymptotic behaviors. In general, black
hole solutions with nonlinear electrodynamics sources have been
extensively analyzed in the current literature, see e.g.
\cite{Breton:2007bza} and references therein. Nonlinear
electrodynamics sources are also good laboratories in order to
construct black-hole solutions with appealing features as for
instance regular black holes~\cite{regularBH}. Moreover, the
nonlinear electrodynamics models exhibit interesting thermodynamics
properties since they satisfy both the zeroth and first laws of
black-hole mechanics~\cite{Rasheed:1997ns}.

The plan of the paper is organized as follows. In the next section,
we consider the Einstein-Gauss-Bonnet action with the nonlinear
electrodynamics source (\ref{nes}). In this case, we derive
electrically charged black-hole solutions and a generalized version
of the Birkhoff's theorem is proved. In the section III, the
properties of the solution are discussed. In the section IV, our
analysis is extended to the full Lovelock action where it is shown
that the metric is given as a solution of a polynomial equation. In
general, this polynomial equation may have no real roots, in which
case the metric solution being purely imaginary is not physically
admissible. Interesting enough, we show that this polynomial
equation always admits at least one real root for the special
election of the Lovelock coefficients given by (\ref{alphapads}).
The summary and the future prospect of the present paper are given
in section V.

\section{Gauss-Bonnet black holes with a nonlinear electrodynamics source}
In this section, we consider the Einstein-Gauss-Bonnet equations
with the nonlinear electrodynamics source (\ref{nes}) in arbitrary
dimensions. The $n$-dimensional action is given by
\begin{align}
S[g_{\mu \nu}, A_{\mu}]&=\int d^nx\sqrt{-g}\biggl[\frac{1}{2\kappa_n^2}(R-2\Lambda+\alpha{L}_{GB}) \biggr] \nonumber \\
&~~~~~~-\beta\int d^nx\sqrt{-g}(F_{\mu\nu}F^{\mu\nu})^q,
\label{action}
\end{align}
where $R$ and $\Lambda$ are $n$-dimensional Ricci scalar and the
cosmological constant, respectively. $F_{\mu\nu}$ is the strength of
the nonlinear electromagnetic field and $q$ is an arbitrary rational
number whose range will be fixed later. Further
$\kappa_n\equiv\sqrt{8\pi G_n}$, where $G_n$ is $n$-dimensional
gravitational constant and $\alpha$ and $\beta$ are the coupling
constants for the Gauss-Bonnet term $L_{GB}$ and the nonlinear
electromagnetic field, respectively. The Gauss-Bonnet term
${L}_{GB}$ is combination of squares of Ricci scalar, Ricci tensor
$R_{\mu\nu}$, and Riemann tensor $R^\mu_{~~\nu\rho\sigma}$ as
\begin{equation}
{L}_{GB} :=
R^2-4R_{\mu\nu}R^{\mu\nu}+R_{\mu\nu\rho\sigma}R^{\mu\nu\rho\sigma}.
\end{equation}
The basic equations following from the action (\ref{action}) are
given by
\begin{align}
{\ma G}^\mu_{~~\nu} &:={G}^\mu_{~~\nu} +\alpha {H}^\mu_{~~\nu} +\Lambda \delta^\mu_{~~\nu}=\kappa_n^2T^\mu_{~~\nu}, \label{beq} \\
0&=\frac{1}{\sqrt{-g}}\partial_{\nu}(\sqrt{-g}F^{\mu\nu}{\cal
F}^{q-1}), \label{max}
\end{align}
where for convenience we have defined ${\cal
F}:=F_{\mu\nu}F^{\mu\nu}$, and where the geometric quantities and
the energy-momentum tensor of the nonlinear electromagnetic field
are defined by
\begin{eqnarray}
{G}_{\mu\nu}&:=&R_{\mu\nu}-{1\over 2}g_{\mu\nu}R,\\
{H}_{\mu\nu}&:=&2\Bigl[RR_{\mu\nu}-2R_{\mu\alpha}R^\alpha_{~\nu}-2R^{\alpha\beta}R_{\mu\alpha\nu\beta} \nonumber \\
&&~~~~+R_{\mu}^{~\alpha\beta\gamma}R_{\nu\alpha\beta\gamma}\Bigr]
-{1\over 2}g_{\mu\nu}{L}_{GB},\label{def-H}\\
T_{\mu\nu}&:=&4\beta\biggl(qF_{\mu\rho}F_{\nu}^{~\rho}{\cal
F}^{q-1}-\frac14 g_{\mu\nu}{\cal F}^q\biggl).
\end{eqnarray}

Now we consider an Ansatz for the spacetime geometry such that the
$n(\ge 5)$-dimensional spacetime $({\ma M}^n, g_{\mu \nu })$ is given as a
warped product of an $(n-2)$-dimensional constant curvature space
$(K^{n-2}, \gamma _{ij})$ and a two-dimensional orbit spacetime
$(M^2, g_{ab})$ under the isometries of $(K^{n-2}, \gamma _{ij})$.
Namely, the line element is given by
\begin{align}
g_{\mu \nu }d x^\mu d x^\nu =g_{ab}(y)d y^ad y^b +{\cal R}^2(y)
\gamma _{ij}(z) d z^id z^j , \label{eq:ansatz}
\end{align}
where $a,b = 0, 1;~i,j = 2, ..., n-1$. Here ${\cal R}$ is a scalar
on $(M^2, g_{ab})$  with ${\cal R}=0$ defining its boundary and
$\gamma_{ij}$ is the unit metric on $(K^{n-2}, \gamma _{ij})$ with
its sectional curvature $k = \pm 1, 0$.

In what follows, we first derive an electrically charged solution
with a particular Ansatz of the form
\begin{eqnarray}
ds^2=-f(r)dt^2+\frac{1}{f(r)}dr^2+r^2\gamma_{ij}dz^idz^j,
\label{metricAnsatz}
\end{eqnarray}
and then we prove that the solution obtained is the unique under the
assumption that $D_{a}{\cal R}$ is not null. Here $D_a$ stands for a
metric compatible linear connection on the manifold $(M^2, g_{ab})$.

\subsection{Electrically charged black-hole solutions}

Here we only consider the electrically charged case, i.e.,
$F_{ij}\equiv 0$, and hence the non-zero components of the
energy-momentum tensor are given by $T^a_{~~b}=\beta(2q-1){\cal
F}^q\delta^a_{~~b}$ and $T^i_{~~j}=-\beta{\cal F}^q \delta^i_{~~j}$.
In this setting, we obtain the following solution for the Ansatz
(\ref{metricAnsatz})
\begin{align}
f(r)&=k+\frac{r^2}{2{\tilde\alpha}}\biggl(1\mp\sqrt{1+4{\tilde\alpha}{\tilde\Lambda}
+\frac{{\tilde\alpha}M}{r^{n-1}}+\frac{{\tilde\alpha}B}{r^{\gamma}}}\biggl),
\label{f} \\
F_{tr}&=\frac{C}{r^{(n-2)/(2q-1)}}, \label{ftr}
\end{align}
for $q \ne 1/2$, where $C$ is a constant and where we have defined
\begin{eqnarray}
B:=\frac{8\kappa_n^2\beta C^{2q}(-2)^q(2q-1)^2}{(n-2)(n-1-2q)},\,\,
\gamma:=\frac{2q(n-2)}{2q-1}. \label{csts}
\end{eqnarray}
The remaining constants appearing in the solution are
${\tilde\Lambda}:=2\Lambda/[(n-1)(n-2)]$, $\tilde{\alpha}:=
(n-3)(n-4)\alpha$, while $M$ stands for an arbitrary constant.

Various comments can be made concerning the solution obtained.
Firstly, this solution reduces to the solutions obtained by Boulware
and Deser, and independently by Wheeler for $C=0$, $k=1$, and
$\Lambda=0$~\cite{bdw}, by Wiltshire for $q=1$, $k=1$, and
$\Lambda=0$~\cite{wiltshire1986}, by Lorenz-Petzold and independently by Cai for $C=0$~\cite{lorenz-petzold1988,cai2002} (the left-hand side of Eq.~(10) in~\cite{lorenz-petzold1988} should be $u^{-2}$), and by Cvetic, Nojiri, and Odintsov for $q=1$~\cite{cno2002}.
Subsequently, it is important to stress that since the only
non-vanishing components of the Maxwell tensor is given by $F_{tr}$,
the Maxwell invariant ${\cal F}=-2(F_{tr})^2$ is negative, and hence
in order to deal with real solutions, the exponent $q$ must be
restricted to be an integer or a rational number with odd
denominator. As a consequence, the singular case of $q=1/2$ is
excluded from the discussion.

It is also clear from the expression of the constant $B$
(\ref{csts}) that the solution given by (\ref{f}) is valid only for
$n\not= 2q+1$. For this particular case, which corresponds to an
exponent $q\in\N$ in odd dimensions $n=2q+1$ (so $q \ge 2$), the solution
reads
\begin{eqnarray}
f(r)=k+\frac{r^2}{2{\bar\alpha}}\biggl(1\mp\sqrt{1+4{\bar\alpha}{\bar\Lambda}
+\frac{{\bar\alpha}M}{r^{2q}}-\frac{{\bar\alpha}\bar{B}\ln(r)}{r^{2q}}}\biggl),\nonumber\\
\label{f2}
\end{eqnarray}
where ${\bar\alpha}:=2(q-1)(2q-3)\alpha$,
${\bar\Lambda}:=\Lambda/[q(2q-1)]$, and $\bar{B}:=8\kappa_n^2\beta
C^{2q}(-2)^q$.

\subsection{Uniqueness}

We now show that the particular solution represented by (\ref{f})--(\ref{f2}) is the
unique solution (up to isometries) under the assumption that
$D_{a}R$ is not null. In what follows, we only consider the case for
which $D_{a}R$ is spacelike since the derivation in the timelike
case is quite analogue.  In the neutral case, i.e., $C=0$, this
generalized Birkhoff's theorem was shown under the same assumption,
i.e., $(D_a R)(D^a R) \ne 0$ in~\cite{wiltshire1986,birkhoff-gb,birkhoff-gb2},
while the complete proof including the null case was given
in~\cite{mn2008}.

In the case where $D_{a}R$ is spacelike, we can set $R$ to be the
radial space coordinate, and in this case the general metric reads
\begin{equation}
ds^2=-N(t,r)g(t,r)dt^2+\frac{1}{g(t,r)}dr^2+r^2\gamma_{ij}dz^idz^j.
\end{equation}
The contravariant-covariant component $(t,r)$ or $(r,t)$ of
Eq.~(\ref{beq}) imply that the metric function $g$ does not depend
on $t$, i.e. $g(t,r)=g(r)$. Subsequently, the combination $({\cal
G}^t_{~~t}-{\cal G}^r_{~~r})-\kappa_n^2(T^t_{~~t}-T^r_{~~r})=0$
gives rise to two possibilities, $N(t,r)=N(t)$ or
\begin{align}
g(r)=k+\frac{r^2}{2{\tilde\alpha}}. \label{special}
\end{align}
In both cases, the nontrivial basic equations (\ref{beq}) are given
by
\begin{eqnarray}
{\ma G}^a_{~~b}&=&\kappa_n^2\beta(2q-1){\cal F}^q\delta^a_{~~b}, \label{b1} \\
{\ma G}^i_{~~j}&=&-\kappa_n^2\beta{\cal F}^q \delta^i_{~~j}, \label{b3}\\
{\cal F}&=&-\frac{2}{N}(F_{tr})^2.
\end{eqnarray}

In the first case, namely $N(t,r)=N(t)$, we can set $N(t) \equiv 1$
without loss of generality. Then, Eq.~(\ref{max}) gives
\begin{align}
0&=\partial_{r}(r^{n-2}F_{tr}(-2F_{tr}^2)^{q-1}), \label{max1} \\
0&=\partial_{t}(r^{n-2}F_{tr}(-2F_{tr}^2)^{q-1}), \label{max2}
\end{align}
from which we deduce that the Maxwell field strength is given by
(\ref{ftr}). Finally, the remaining metric function $g(r)$ is given
by $g(r)=f(r)$, where $f(r)$ is expressed as Eq.~(\ref{f}) and
(\ref{f2}) for $n\ne 2q+1$ and $n=2q+1$, respectively.

We now analyze the remaining option $g(r)=k+r^2/(2{\tilde\alpha})$.
In this case, considering the equation ${\ma G}^t_{~~t}+(2q-1){\ma
G}^2_{~~2}=0$ given from Eqs.~(\ref{b1}) and (\ref{b3}), we obtain
that ${\tilde\Lambda}=-1/(4\tilde{\alpha})$. Through the basic
equations (\ref{b1}) and (\ref{b3}) with
$g(r)=k+r^2/(2{\tilde\alpha})$ and
${\tilde\Lambda}=-1/(4\tilde{\alpha})$ imply that the system is
vacuum, i.e., $T^\mu_{~~\nu}\equiv 0$, and $N(t,r)$ is arbitrary.
This exceptional vacuum (non-static) solution under the special combination between the coupling constants $\alpha$ and $\Lambda$ was first found in~\cite{birkhoff-gb}.

Here we have shown the uniqueness of our solution (\ref{f})--(\ref{f2}) under the assumption that $D_{a}R$ is not null.
For the null case, on the other hand, there must be the
Nariai-Bertotti-Robinson type solution~\cite{nbr} as in the case with or without the Maxwell field
in general relativity~\cite{nbr-gr} and in the Einstein-Gauss-Bonnet
gravity~\cite{nbr-gb,nbr-gb2}.

\subsection{Energy conditions}

Before analyzing the properties of the solutions obtained in
(\ref{f}) and (\ref{f2}), we discuss the energy conditions for the
nonlinear electromagnetic field. For the energy momentum tensor
written in the diagonal form as
$T^\mu_{~~\nu}=\mbox{diag}(-\mu,p_{\rm r},p_{\rm t},p_{\rm
t},\cdots)$, the weak energy condition (WEC) implies $\mu \ge 0$,
$p_{\rm r}+\mu \ge 0$, and $p_{\rm t}+\mu \ge 0$, while the dominant
energy condition (DEC) implies $\mu \ge 0$, $-\mu \le p_{\rm r} \le
\mu$, and $-\mu \le p_{\rm t} \le \mu$~\cite{he}. The physical
interpretations of $\mu$, $p_{\rm r}$, and $p_{\rm t}$ are energy
density, radial pressure, and the tangential pressure, respectively.
The WEC assures that a timelike observer measures the non-negative
energy density. The DEC assures in addition that the energy flux is
a future-directed causal vector. The DEC implies the WEC, but the
converse is not true.

In our case, the corresponding $\mu$, $p_{\rm r}$, and $p_{\rm t}$
are respectively given by
\begin{eqnarray}
\mu&=&-\beta(2q-1){\cal F}^q, \label{mu} \\
p_{\rm r}&=&\beta(2q-1){\cal F}^q, \label{pr} \\
p_{\rm t}&=&-\beta{\cal F}^q, \label{pt}\\
{\cal F}&=&-2(F_{tr})^2.
\end{eqnarray}
It is noted again that the exponent $q$ must be restricted to be an
integer or a rational number with odd denominator in order to deal
with real solutions, so that $q=1/2$ is excluded.

First we consider the condition $\mu \ge 0$, which determines the
sign of $\beta$ depending on the range of the exponent $q$,
\begin{equation}
\left\lbrace
\begin{array}{l}
\mbox{sgn}(\beta)=-(-1)^q\qquad\mbox{for}\quad q>1/2,\\
\mbox{sgn}(\beta)=(-1)^q\qquad\quad\mbox{for}\quad q<1/2.
\end{array}
\right. \label{sgn}
\end{equation}

Then, both WEC and DEC are satisfied for $q \le 0$ or $q \ge 1$. The
WEC is satisfied but the DEC is violated for $1/2<q<1$. On the other
hand, both WEC and DEC are violated for $0<q <1/2$.

We also consider the strong energy condition (SEC) which implies
$p_{\rm r}+\mu \ge 0$, $p_{\rm t}+\mu \ge 0$, and $\mu+p_{\rm
r}+(n-2)p_{\rm t} \ge 0$. It is noted that the SEC is independent
from either WEC or DEC. Independent of the sign of $\beta$, the SEC
is satisfied for $q>1/2$ with $\mu >0$ or $0\le q<1/2$ with $\mu <0$,
otherwise it is violated. The result obtained in this subsection is
summarized in table~\ref{table}.
\begin{table}[h]
\begin{center}
\caption{\label{table} Consistency for the nonlinear electromagnetic
field with the energy conditions under the condition (\ref{sgn})
corresponding to $\mu>0$. For the real solutions, the exponent $q$
must be an integer or a rational number with odd denominator. We
note that the strong energy condition is satisfied even for $0\le q<1/2$
if $\mu <0$ holds.}
\begin{tabular}{|c||c|c|c|c|c|}
\hline \hline
  & ~~~~$q \le 0$~~~~ & $0<q<1/2$ & $1/2 < q<1$ & ~~~~$1\le q$~~~~  \\\hline
WEC & Yes & No & Yes & Yes \\ \hline DEC & Yes & No & No & Yes  \\
\hline
SEC & No & No & Yes & Yes \\
\hline \hline
\end{tabular}
\end{center}
\end{table}

To conclude the study of the energy condition, we would like to
stress that the excluded region $0<q<1/2$ where none of the energies
conditions are satisfied is also ruled out by the following
argument. Since we are interested in finding solutions with event
horizons that should hide the eventual singularities, solutions
having singularities at infinity will be ruled out and only
curvature singularities surrounded by an event horizon will be
allowed. For $q\in]0,1/2[$, the scalar curvature associated to the
solution (\ref{f}) diverges at infinity or 
the metric may be complex at infinity depending on the parameters.

\section{Properties of the solution}

In this section, we analyze the solutions obtained (\ref{f}) and
(\ref{f2}). There are two families of solutions corresponding to the
sign in front of the square root in Eq.~(\ref{f}) or (\ref{f2}),
stemming from the quadratic curvature terms in the action. The
solution with the upper sign, that we call the GR branch, has a
general relativistic (GR) limit as $\alpha \to 0$ given by
\begin{align}
f(r)&=k-{\tilde\Lambda}r^2-\frac{M}{4r^{n-3}}-\frac{B}{4r^{\gamma-2}},
\label{f-gr}\\
f(r)&=k-{\bar\Lambda}r^2-\frac{M}{4r^{2q-2}}+\frac{\bar{B}\ln(r)}{r^{2q-2}}
\label{f2-gr}
\end{align}
for $n \ne 2q+1$ and for $n=2q+1$ respectively. This is a
generalization of the solution obtained in \cite{hm2008} for $k=1$
and $\Lambda=0$. In contrary, the other branch, i.e. the lower signs
in (\ref{f}) and (\ref{f2}), that we call the Gauss-Bonnet branch,
does not have the GR limit.

Setting $\tilde{\Lambda}=\bar{\Lambda}=M=C=0$ in (\ref{f}) or
(\ref{f2}), the possible vacua differ drastically from one case to
the other. Indeed, in the GR branch, the metric will reduce to that
of Minkowski while for the Gauss-Bonnet branch, the metric becomes
that of (A)dS with an effective cosmological constant that goes like
$-(1/\alpha)$. Indeed, in this case a small coupling constant
$\alpha$ will correspond to a huge effective cosmological constant.

We now turn to the crucial question about the singularities and the
existence of event horizons. In order to achieve this task correctly
and because of the presence of many parameters in the metric
solution (\ref{f}), we put several conditions on the parameters.
First we assume $1+4{\tilde\alpha}{\tilde \Lambda} \ge 0$ and
$1+4{\bar \alpha}{\bar \Lambda} \ge 0$ which ensure the existence of
the maximally symmetric solutions. We also assume the weak energy
condition for the nonlinear electromagnetic field, under which $q
\le 0$ or $q>1/2$ is satisfied and $\gamma$ is non-negative.

Under the reasonable assumptions listed above, the parameter space
of the solution may be classified into several cases depending on
the fall-off rate of the electromagnetic term against the
gravitational term. The first case corresponding to $\gamma>n-1$ is
similar to the standard Maxwell case and will be achieved for the
exponent $q\in]1/2,(n-1)/2[$. On the other hand, the option
$\gamma<n-1$ can also be considered with $q \le 0$ or $q>(n-1)/2$
while the case $\gamma=(n-1)$ will correspond to the logarithmic
metric (\ref{f2}).

In a generic way, the solution may have two possible singularities
that are the usual $r=0$ and also a branch singularity at $r=r_{\rm
b}(>0)$, where the argument of the square-root piece of the metric
solution (\ref{f}) or (\ref{f2}) vanishes. For $r<r_{\rm b}$, the
metric becomes complex.

In order to clarify the existence condition for the branch
singularity, we write the function $f(r)$ as
\begin{widetext}
\begin{eqnarray}
f(r)&=&k+\frac{r^2}{2{\tilde\alpha}}\biggl(1\mp\sqrt{1+4{\tilde\alpha}{\tilde\Lambda}
+\frac{{\tilde\alpha}M}{r^{n-1}}-\frac{8{\tilde\alpha}\kappa_n^2(2q-1)\mu}{(n-1)(n-1-2q)}}\biggl), \label{f-sp} \\
f(r)&=&k+\frac{r^2}{2{\bar\alpha}}\biggl(1\mp\sqrt{1+4{\bar\alpha}{\bar\Lambda}
+\frac{{\bar\alpha}M}{r^{2q}}+\frac{8\kappa_n^2{\bar\alpha}\mu\ln(r)}{2q-1}}\biggl)
\label{f2-sp}
\end{eqnarray}
\end{widetext}
for $n \ne 2q+1$ and for $n=2q+1$ respectively, where $\mu(r)$ is the
energy density of the nonlinear electromagnetic field (\ref{mu}).

For $q\in]1/2,(n-1)/2[$ corresponding to $\gamma>n-1$, the
electromagnetic term dominates inside the square-root for $r \to 0$.
As a result, the branch singularity exists for ${\tilde\alpha}>0$.
On the other hand, for $q \le 0$ or $q>(n-1)/2$ corresponding to
$\gamma<n-1$, the gravitational term dominates for $r \to 0$ and the
branch singularity exists for ${\tilde\alpha}M<0$. Finally, in the
case of $q=(n-1)/2$ corresponding to the logarithmic metric
(\ref{f2-sp}), the electromagnetic term dominates for $r \to 0$, so
that the branch singularity exists for ${\bar\alpha}>0$ since
$2q-1>0$ is satisfied.

As a consequence of the branch singularity, the event horizon given
by the positive real root of the algebraic equation $f(r_{\rm h})=0$
must satisfy an inequality $r_{\rm h}>\mbox{max}(0,r_{\rm b})$. The
location of horizon $r_{\rm h}$ is a root of the following
polynomial
\begin{align}
p(r):=4k^2\tilde{\alpha}r^{\gamma-4}+4kr^{\gamma-2}-4\tilde{\Lambda}r^{\gamma}-Mr^{\gamma+1-n}-B=0
\end{align}
for $\gamma>n-1$, while for $\gamma<n-1$, the polynomial reads
\begin{align}
p(r)&:=4k^2\tilde{\alpha}r^{n-5}+4kr^{n-3} \nonumber \\
&~~~~~~~~~-4\tilde{\Lambda}r^{n-1}-Br^{n-\gamma-1}-M=0.
\end{align}
For the logarithmic case, i.e., $\gamma=n-1$, the horizon $r_{\rm
h}$ is the solution of
\begin{align}
4k^2\bar{\alpha}r^{2q-4}+4kr^{2q-2}-4\bar{\Lambda}r^{2q}+\bar{B}\ln(r)-M=0,
\end{align}
which is not a polynomial. Moreover, in all the cases, the roots
must satisfy the condition $\mp[k+r_{\rm h}^2/(2\tilde{\alpha})]\leq
0$, where the upper and the lower signs in the left-hand side
correspond to the GR and the Gauss-Bonnet branches, respectively.
This extra condition ensures the equivalence between the roots of
the polynomial $p(r)$ and those of the metric function $f(r)$. A
full analysis of the existence condition for the event horizon will
certainly be interesting but it is rendered long by the
presence of so many parameters. Hence, in order to gain in clarity
we avoid  a more detailed discussion.

In the case of $q=1$, the global structure of the solution was fully
investigated in~\cite{tm2005b}. (See~\cite{tm2005} for the neutral
case.) In fact, the number of the horizons, structure of the
singularity, and asymptotic behavior at infinity, sharply depend on
the parameters in the solution.

\section{Extension to Lovelock gravity}

The extension of the analysis in the previous section to the more
general Lovelock gravity is an interesting subject by itself. In
this objective, we shall consider the Lovelock gravity
(\ref{LovLag}) with the nonlinear electrodynamics source (\ref{nes})
in arbitrary dimensions and look for particular solutions. The
Ansatz for the geometry we shall consider is the same that in the
Gauss-Bonnet case (\ref{metricAnsatz}), and we shall also restrict
the nonlinear electromagnetic field to be a purely radial one. As in
the Gauss-Bonnet case, this Ansatz will restrict the exponent $q$ to
be given as a rational number with odd denominator.

In this analysis, we opt for the Hamiltonian formalism that provides
an easy way to write down the field equations and integrate them. In
order to achieve this task, we first write the reduced Lovelock
Hamiltonian \cite{btz1994},
\begin{align}
{\cal
H}^{L}=-(n-2)!\sqrt{\frac{\gamma}{f}}\frac{d}{dr}\left[r^{n-1}\sum_{p=0}^{[\frac{n-1}{2}]}\alpha_p(n-2p)
\left(\frac{k-f}{r^2}\right)^p\right],
\end{align}
where $\gamma$ is the determinant of $\gamma_{ij}$. In an analogue
way, the reduced nonlinear electromagnetic Hamiltonian is given by
\begin{align}
{\cal
H}^{e}=-\beta\sqrt{\frac{\gamma}{f}}\,\frac{(2q-1)\,(-2)^{\frac{q}{2q-1}}\,{\cal
P}^{\frac{2q}{2q-1}}}{(4q\beta)^{\frac{2q}{2q-1}}\,r^{\frac{n-2}{2q-1}}},
\end{align}
where ${\cal P}:=4\beta q {\cal F}^{q-1}r^{n-2}F_{tr}$ is the
rescaled radial momentum which is constant by virtue of the Gauss
law. Defining a function $H(r)$ such that $f(r)=k-r^2H(r)$, the
constraint becomes a first-order equation given by
\begin{align}
&\frac{d}{dr}\left[r^{n-1}\sum_{p=0}^{[\frac{n-1}{2}]}\alpha_p(n-2p)H^p\right] \nonumber \\
&~~~~~~~~~~=\frac{\beta}{(n-2)!}\frac{(2q-1)\,(-2)^{\frac{q}{2q-1}}\,{\cal
P}^{\frac{2q}{2q-1}}}{(4q\beta)^{\frac{2q}{2q-1}}\,r^{\frac{n-2}{2q-1}}},
\end{align}
whose straightforward integration yields
\begin{widetext}
\begin{align}
\sum_{p=0}^{[\frac{n-1}{2}]}\alpha_p(n-2p)
H^p=\frac{C_1}{r^{n-1}}+\frac{\beta(2q-1)^2\,(-2)^{\frac{q}{2q-1}}\,{\cal
P}^{\frac{2q}{2q-1}}}{(n-2)!(4q\beta)^{\frac{2q}{2q-1}}\,(n-1-2q)\,r^{\frac{2q(n-2)}{2q-1}}},\qquad
\mbox{for}\quad n\not=2q+1 \label{eqa-2} \\ 
\sum_{p=0}^{[\frac{n-1}{2}]}\alpha_p(n-2p)
H^p=\frac{C_1}{r^{n-1}}-\frac{\beta(2q-1)(-2)^{\frac{q}{2q-1}}\,
{\cal P}^{\frac{q}{2q-1}}\,\ln
r}{(2q-1)!\,(4q\beta)^{\frac{q}{2q-1}}\,r^{2q}},\qquad
\mbox{for}\quad n=2q+1, \label{eqa}
\end{align}
\end{widetext}
where $C_1$ is an integration constant in both cases. In both cases,
the electric field is given by the same expression as in the
Gauss-Bonnet case (\ref{ftr}).

\subsection{Dimensionally continued gravity}
\label{dc}

In principle, one may find up to $[\frac{n-1}{2}]$ real roots and in
dimensions $n=4m+3$ and $4m+4$ with an integer $m(\geq 1)$, these
equations will always admit at least one real root. However, it is
interesting to observe that an enormous simplification occurs in
these equations as the Lovelock coefficients $\alpha_p$ takes the
particular values (\ref{alphapads}) that convert the Lovelock action
into a Chern-Simons gauge theory in odd dimensions. Indeed, in this
case, and for odd as well as even dimensions, the left-hand sides of
the equations (\ref{eqa-2}) and (\ref{eqa}) become the Newton binomial expression,
\begin{eqnarray}
\sum_{p=0}^{[\frac{n-1}{2}]}\alpha_p(n-2p) H^p \equiv
\Big(1+H\Big)^{[\frac{n-1}{2}]}. \label{Newton}
\end{eqnarray}
Consequently, in both cases, the function $H$ can be determined
explicitly. The metric solution $f$ can be written in odd dimension
by
\begin{eqnarray}
f(r)=k+r^2-\Big(C_1+\frac{C_2}{r^{\frac{n-1-2q}{2q-1}}}\Big)^{\frac{2}{n-1}},
\label{fo}
\end{eqnarray}
while the expression in even dimension is giving by
\begin{eqnarray}
f(r)=k+r^2-\frac{1}{r^{\frac{2}{n-2}}}\Big(C_1+\frac{C_2}{r^{\frac{n-1-2q}{2q-1}}}\Big)^{\frac{2}{n-2}},
\label{fe}
\end{eqnarray}
where $C_2$ stands in both cases for
\begin{align}
C_2:=\frac{\beta(2q-1)^2\,(-2)^{\frac{q}{2q-1}}\,{\cal
P}^{\frac{2q}{2q-1}}}{(n-2)!(4q\beta)^{\frac{2q}{2q-1}}\,(n-1-2q)}.
\end{align}
This solution is a generalization of the solution obtained by
Ba\~nados, Teitelboim, and Zanelli for $k=1$ and $q=1$
in~\cite{btz1994} and by Cai and Soh for $q=1$ in~\cite{cs1999}. In
odd dimensions, this solution is the higher-dimensional counterpart
with a nonlinear electromagnetic charge of the so-called BTZ black
hole in three dimensions~\cite{btz}. It is very appealing that for
the special election of the Lovelock coefficients, the metric
function can be integrated easily in odd as well as even dimensions.

\subsection{Properties of the solution}

Now let us discuss the properties of the solution (\ref{fo}) and
(\ref{fe}) under the weak energy condition. The constant $C_2$ can
be written in terms of the energy density of the nonlinear
electromagnetic field (\ref{mu}) as
\begin{align}
C_2=-\frac{(2q-1)\mu r^{\frac{2q(n-2)}{2q-1}}}{(n-2)!(n-1-2q)}.
\end{align}
Therefore, under the weak energy condition, $C_2<0$ for
$q\in]1/2,(n-1)/2[$ while $C_2>0$ for $q \le 0$ or $q>(n-1)/2$.

In the case for which the dimension is expressed as $n=4m+1$ or
$n=4m+2$, where $m(\geq 1)$ is an integer, the exponent in the
binomial expression (\ref{Newton}) is even and hence the equations
(\ref{eqa}) have two branches of solutions
\begin{align}
f(r)=k+r^2\mp\Big(C_1+\frac{C_2}{r^{\frac{4m-2q}{2q-1}}}\Big)^{\frac{1}{2m}},\quad
\mbox{for}\quad n=4m+1,\label{j1}\\
f(r)=k+r^2\mp\Big(\frac{C_1}{r}+\frac{C_2}{r^{\frac{4m}{2q-1}}}\Big)^{\frac{1}{2m}},\quad
\mbox{for}\quad n=4m+2 \label{j2}.
\end{align}
Note that the Einstein-Gauss-Bonnet solution derived previously
(\ref{f}) in five dimensions with the special election
$\tilde{\alpha}=-1/(4\tilde{\Lambda})$ reduces to the first
expression (\ref{j1}) with $m=1$. This is not surprising since the
condition $\tilde{\alpha}=-1/(4\tilde{\Lambda})$ is nothing but the
Chern-Simons limit of the Gauss-Bonnet theory in five dimensions.

For the solutions (\ref{j1}-\ref{j2}), we have to care about the
possible branch singularities. In these cases, the gravitational
term dominates the nonlinear electromagnetic term at infinity for
$q\in]1/2,(n-1)/2[$, so that $C_1$ must be positive in order that
the metric is real at infinity. On the other hand, there exists a
branch singularity since $C_2<0$ is required by the weak energy
condition. For $q \le 0$ or $q>(n-1)/2$, the nonlinear
electromagnetic term dominates the gravitational term at infinity
and the metric is real because the weak energy condition requires
$C_2>0$. In this case, there exists a branch singularity for $C_1<0$
and a central singularity for $C_1>0$.

In the case with $n=4m+3$ or $n=4m+4$, on the other hand, both $C_1$
and $C_2$ may be negative since we may rewrite (\ref{fo}) and
(\ref{fe}) as
\begin{eqnarray}
f(r)=k+r^2+\Big(-C_1-\frac{C_2}{r^{\frac{n-1-2q}{2q-1}}}\Big)^{\frac{2}{n-1}},
\end{eqnarray}
and
\begin{eqnarray}
f(r)=k+r^2+\frac{1}{r^{\frac{2}{n-2}}}\Big(-C_1-\frac{C_2}{r^{\frac{n-1-2q}{2q-1}}}\Big)^{\frac{2}{n-2}},
\end{eqnarray}
respectively. The branch singularity exists only for $C_1C_2<0$.
Unlike the case of $n=4m+1$ or $n=4m+2$, there is no region where
the metric becomes complex even if there exists a branch
singularity. Under the weak energy condition, the branch singularity
exists for $q\in]1/2,(n-1)/2[$ with $C_1>0$ and for $q \le 0$ or
$q>(n-1)/2$ with $C_1<0$.

In the solution (\ref{fo}) and (\ref{fe}), the fall-off rate to
infinity is slower than the standard one. Even under the weak energy
condition, the electromagnetic term diverges at infinity for $q <
1/2$ or $q>(n-1)/2$. However, it is shown that the divergence is
faster than $r^2$ only for $0<q<1/2$ both in odd and even
dimensions, which is ruled out by the weak energy condition. As a
result, the regular infinity is assured by the weak energy
condition. This slow fall-off phenomenon was first pointed out in
the study of the static black holes with and without the Maxwell
field in the class of Lovelock gravity admitting a unique (A)dS
vacuum~\cite{ctz2000}. Recently, this phenomenon was shown to be
universal for any matter field satisfying the dominant energy
condition in Einstein-Gauss-Bonnet gravity with
$1+4{\tilde\alpha}{\tilde\Lambda}=0$ and
$\alpha>0$~\cite{maeda2008}.

We finally end this section with some speculation concerning a
possible Birkhoff's theorem. In Lovelock gravity, the generalized
Birkhoff's theorem has been proven under the same assumption of the present paper in the vacuum case and for the standard Maxwell case~\cite{zegers2005,df2005}. Because of the similarity in the
treatment, we may envisage that our charged solution is the unique
electrically charged solution within the nonlinear source considered
here.

\section{Summary and further prospects}

In the present paper, we obtained electrically charged black-hole
solutions in Einstein-Gauss-Bonnet gravity with a nonlinear source
given as an arbitrary exponent $q$ of the Maxwell invariant. We have
considered the class of the $n(\ge 5)$-dimensional spacetime given
as a warped product ${\ma M}^{2} \times {\ma K}^{n-2}$. The generic
solution is shown to have two branches and only one of them has a GR
limit. For an integer value of $q$ with a dimension $n=2q+1$, the
metric solution involves a logarithmic dependence and as in the
generic case, the solution presents as well two different branches.
We show that these solutions are the unique electrically charged
solution in the case where the orbit of the warp factor on ${\ma
K}^{n-2}$ is non-null. 

We find that an intriguing slow fall-off to the spacelike infinity
is possible even under the dominant energy condition. This slow
fall-off was shown to be universal for any matter field satisfying
the dominant energy condition in the special case of
$1+4{\tilde\alpha}{\tilde\Lambda}=0$ and $\alpha>0$, which
corresponds to the Chern-Simons gravity in five
dimensions~\cite{ctz2000,maeda2008}. Our solution is an example
exhibiting the slow fall-off with generic coupling constant. We have
analyzed the properties of the solutions emphasizing the study on
the branch singularities.

We have also derived charged black hole solutions for the full
Lovelock gravity. In this case, the metric function is obtained
implicitly as a solution of a polynomial equation. We have pointed
out that for a very precise combination between the coupling
constants, which converts the Lovelock action into a Chern-Simons
gauge theory in odd dimensions, the metric function is obtained in a
closed form both in odd and even dimensions. It has been shown that
a branch singularity appears under the weak energy condition
depending on the parameters. Unlike the Gauss-Bonnet case, the
metric does not become complex for $n=4m+3$ and $n=4m+4$ with an
integer $m(\ge 1)$ even if there is a branch singularity. The slow
fall-off to the regular infinity is a generic property under the
weak energy condition both in odd and even dimensions.

This slow fall-off has recently attracted much attention ~\cite{slow} for
theories of AdS gravity coupled to a scalar field with mass at or
slightly above the Breitenlohner-Freedman bound~\cite{bf}. These
theories  admit a
large class of asymptotically AdS spacetimes with slower fall-off
conditions than the standard ones.

The dynamical stability of the black-hole solution is an important
problem. In the standard Einstein-Maxwell case, the stability
analysis has been carried out in four dimensions~\cite{BHstability4}
as well as in higher dimensions~\cite{BHstability}, see Table I in
~\cite{kodama2007} for the summary of the analytic results. In the
Einstein-Gauss-Bonnet gravity, the analysis has been done only in
the neutral case~\cite{GBBH-stability}, while there is no result for
higher-order Lovelock gravity. In the case of the nonlinear
electromagnetic field, the stability analysis has not been done even
in four dimensions. In this context, the asymptotic slow fall-off
would be important because it could affect the boundary conditions
for the perturbations.

The black-hole thermodynamics is another interesting subject. In
Einstein-Gauss-Bonnet gravity, this subject have been intensively
investigated with or without the Maxwell
charge~\cite{GBBH-thermo,cno2002,cai2002,tm2005}. The extension to
the full Lovelock gravity is also well studied~\cite{btz1994,lovelock-thermo,cs1999}. 
In this context, the slow
fall-off to the spacelike infinity becomes important. Under the
standard fall-off condition, the higher-dimensional ADM mass is
available as the global mass in the asymptotically flat case~\cite{mp1986}. In the
asymptotically (A)dS spacetime, several definitions of the global
mass have been proposed in Einstein-Gauss-Bonnet
gravity~\cite{padilla2003,deser2002,ok2005}. However, the slow fall-off
means that they are diverging at infinity. 
In order to discuss the
thermodynamical properties of black holes correctly, one should
first reformulate the global mass in order to give a finite value
under the slower fall-off condition in this special case. This
problem has been investigated in Chern-Simons gravity~\cite{cs-mass}
and in the theory admitting a unique (A)dS vacuum~\cite{ctz2000}.
(See also~\cite{dko2003,br2008}.)

Other aspects to explore are the extensions of the solutions
presented here in more general context. For example, it will be
interesting to explore the possible dilatonic solutions in this
set-up or the existence of magnetically charged solutions.
These prospects presented here are left for possible future
investigations.

\acknowledgments 
HM would like to thank Julio Oliva for discussions and useful comments and Dieter Lorenz-Petzold for references on the Nariai-Bertotti-Robinson type solution.
MH would like to thank Christos Charmousis for useful discussions.
This work was partially funded by the following Fondecyt grants:
1071125 (HM); 1051084, 1060831 (MH); and  1051056, 1061291, 1085322
(CM). The Centro de Estudios Cient\'{\i}ficos (CECS) is funded by
the Chilean Government through the Millennium Science Initiative and
the Centers of Excellence Base Financing Program of Conicyt. CECS is
also supported by a group of private companies which at present
includes Antofagasta Minerals, Arauco, Empresas CMPC, Indura,
Naviera Ultragas, and Telef\'{o}nica del Sur. CIN is funded by
Conicyt and the Gobierno Regional de Los R\'{\i}os. Finally, we
thank the participants of\textit{ CECS Theoretical Physics Group
Wokshop 2008: ``Puerto Octay"}, where this work started, for
enlightening discussions.



\end{document}